\documentclass[12pt]{article}
\pdfoutput=1 
\input epsf.sty
\usepackage{pdfpages}
\newcommand{\be}{\begin{equation}}
\newcommand{\ee}{\end{equation}}
\newcommand{\ben}{\begin{eqnarray}\displaystyle}
\newcommand{\een}{\end{eqnarray}}

\newcommand{\eg}{elliptic genus }
\usepackage{amssymb}
\usepackage{amsmath}
\usepackage{simplewick}
\usepackage[all]{xy}

\def\be{\begin{equation}}
\def\ee{\end{equation}}
\def\ba{\begin{align}}
\def\ea{\end{align}}

 \input{epsf}
 \usepackage{epsfig}

\def\sqr#1#2{{\vcenter{\vbox{\hrule height.#2pt
         \hbox{\vrule width.#2pt height#1pt \kern#1pt
            \vrule width.#2pt}
         \hrule height.#2pt}}}}

\begin{document}

\baselineskip=18pt

\title{\Large{\bf Lessons on Black Holes from 
the 
Elliptic Genus}}
\author{Amit Giveon$^{a}$, Nissan Itzhaki$^{b}$ and Jan Troost$^{c}$  }
\date{}
\maketitle
\begin{center}
$^{a}$
Racah Institute of Physics, The Hebrew University, Jerusalem, 91904, Israel\\
$^{b}$ Physics Department, Tel-Aviv University,
Ramat-Aviv, 69978, Israel\\
   $^{c}$Laboratoire de Physique Th\'eorique \\
 Unit\'e Mixte du CNRS et
     de l'\'Ecole Normale Sup\'erieure \\ associ\'ee \`a l'Universit\'e Pierre et
     Marie Curie 6
\\ \'Ecole Normale Sup\'erieure \\
   $24$ Rue Lhomond Paris $75005$, France
\end{center}

 \begin{abstract}
We further study the elliptic genus of the cigar  $SL(2,\mathbb{R})_k/U(1)$ coset
superconformal field theory.
We find that, even in the
small curvature, infinite level limit,
there are  holomorphic and non-holomorphic parts
 that are due to the discrete states and a mismatch in the spectral densities
of the continuum, respectively.
 The mismatch in the continuum is universal,
 in the sense that it is fully determined by the asymptotic cylindrical topology of the cigar's throat.
 Since modularity of the \eg requires both the holomorphic and non-holomorphic parts,
  the holomorphic term is universal as well.
 The contribution of the discrete states is thus present even for perturbative strings
 propagating in the background of {\it large} Schwarzschild black holes.
 We argue that the discrete states live at a stringy distance from the tip of the cigar both
 from the
conformal field theory
wave-function analysis and from a holonomy space perspective.
 Thus, the way string theory takes care of its self-consistency
seems to  have important consequences for the physics near horizons,
 even for parametrically large black holes.
\end{abstract}

\newpage

\tableofcontents

\newpage

\section{Introduction }

Recently \cite{Troost:2010ud, Eguchi:2010cb, Ashok:2011cy},
progress was made in the understanding of the \eg associated with the cigar geometry
obtained in the $SL(2, \mathbb{R})_k/U(1)$ coset superconformal field theory.
In this paper, we take advantage of these developments and address a question that was raised in the last year
\cite{Giveon:2012kp,Giveon:2013ica}:
Is the tip of the cigar geometry special in the small curvature limit?

{}From a general relativity perspective, it is hard to see how this
can come about, since the curvature and dilaton slope at the tip
vanish in the $k\to\infty$ limit.  Nevertheless, motivated by \cite{Kutasov:2005rr}, it was argued
 \cite{Giveon:2012kp,Giveon:2013ica}
that even in the large $k$ limit there are stringy degrees of freedom
with wave-functions that are supported at the tip of the cigar and
that fall off exponentially fast at distances larger than the string scale. (See also
\cite{Mertens:2013zya}, where the relevance of these degrees of freedom
to string theory thermodynamics near the Hagedorn transition was further investigated.)
These stringy degrees of freedom are in the discrete representations
of the $SL(2, \mathbb{R})_k/U(1)$ model.

The \eg is an extremely useful tool to carefully address the existence of these states.
The reason is that, unlike the partition function, it is
free of divergences associated with the infinite volume of the cigar at any finite
level $k$ and it
gets contributions from the discrete states. Thus, the large $k$ limit of the \eg
reveals unambiguously the fate of the discrete states.

In this paper, we take the large level limit of the \eg in two complementary
ways.  In section \ref{largelevel}, we use the Hamiltonian form and
find that the holomorphic piece due to the discrete states is finite
even at $k\to\infty$.  The non-holomorphic piece, due to a mismatch in
spectral densities between the bosonic and fermionic continuum, is
also finite.  Aspects of this mismatch are reviewed in section
\ref{generalities}. Modularity of the \eg requires the existence of
both terms. We conclude that the discrete states survive the
large $k$ limit.

Following \cite{Dijkgraaf:1991ba,Ashok:2013kk},
we analyze the target-space wave-functions associated with
the discrete states using the $SL(2, \mathbb{R})_k/U(1)$ Laplacian eigenvalues and find that,
indeed, in the large $k$ limit they are located  at the tip of the cigar.
In section \ref{universality}, we argue that the answer obtained for the \eg in the large $k$ limit is universal.
That is, provided the non-holomorphic part of the elliptic genus
is non-trivial,  we expect to find similar states also for cigar geometries associated with other black holes,
including Schwarzschild in four dimensions,
in the limit where their mass is parametrically large.

In section \ref{saddle},
we use the Lagrangian approach to study  the large $k$ limit of the elliptic genus.
We show that the saddle point approximation is particularly efficient.
It also suggests a novel UV/IR relation between the cigar target space and the holonomy space,
discussed in section \ref{uvir}.
This UV/IR relation can be used to find the location of states in target space knowing
the size of the region in holonomy space that supports them. Using this approach,
we confirm that the holomorphic term in the \eg comes from states that live at the tip of the cigar.
In section \ref{trumpet},
we refine our discussion of universality via inspection of the elliptic genus of the trumpet
conformal field theory.
Finally, we conclude in section \ref{conclusions}, and present useful formulas in the appendix.

\section{The physics of non-compact elliptic genera}
\label{generalities}

In this section, we present generic features of non-compact elliptic
genera.  We focus on the physical meaning of the holomorphic
anomaly in the non-compact elliptic genus. We start by illustrating that the origin of the anomaly can
be traced to the calculation of a Witten index \cite{Wittenindex} in
supersymmetric quantum mechanics with a spectrum that contains both
discrete states and a continuum (see e.g. \cite{Akhoury:1984pt}). Then
 we
show that those properties are inherited by a two-dimensional
generalization of the Witten index, the elliptic genus.\footnote{The
  properties we present are discussed in references
  \cite{Troost:2010ud,Ashok:2011cy,Ashok:2013kk} -- it may be
 convenient to gather them.} We review how these features apply to the $SL(2, \mathbb{R})_k/U(1)$ model.
We thus  set the stage for a detailed study of the \eg of
the $SL(2, \mathbb{R})_k/U(1)$ coset superconformal field theory at
large level in section \ref{largelevel}.
\subsection{The Witten index and the continuum}
\label{Wittenind}
Consider a supersymmetric quantum mechanics with at least one supersymmetry and
a fermion number operator $F$ that takes value $0$ on bosons and $1$ on fermions.
We study the weighted trace $Z$,
\begin{eqnarray}
Z &=& \mbox{Tr} \, (-1)^F e^{- \beta H}  \, ,
\end{eqnarray}
where $H$ is the Hamiltonian, and $\beta$ the inverse temperature of
the system. When the spectrum is discrete, the weighted trace is equal
to an index, the Witten index \cite{Wittenindex}, that counts the
number of bosonic minus the number of fermionic ground states.

When the spectrum of the supersymmetric quantum mechanics contains a continuum, the weighted trace will
obtain a contribution from the integral over the continuum, with a measure given by the difference in spectral
densities between bosons and fermions (see e.g. \cite{Akhoury:1984pt}),
\begin{eqnarray}
 \mbox{Tr} \, (-1)^F e^{- \beta H}
 &=& N_{bos}^{ground}-N_{ferm}^{ground} + \int_{cont} d E (\rho_{bos}(E)-\rho_{ferm}(E)) e^{- \beta E} \, .
 \end{eqnarray}
 The difference in spectral densities is generically non-trivial, and
 will contribute a temperature dependent term to the weighted
 trace. Supersymmetry relates the spectral densities.

 Consider for concreteness a non-relativistic supersymmetric quantum
 mechanics defined on a real line (or half-line) with a scattering wall
 on one side. Then the difference in spectral densities will be given
 by a derivative with respect to continuous momentum, $s$,
 of the ratio of the bosonic
 and fermionic phase shifts,
 \begin{eqnarray}
 \Delta\rho\equiv\rho_{bos}-\rho_{ferm} &=& \frac{1}{2 \pi i}
\frac{d}{ds} \log \frac{R_{bos}}{R_{ferm}} \, .
 \end{eqnarray}
 The phase shifts are the logarithms of the corresponding reflection
 amplitudes $R_{bos/ferm}$.  Since the asymptotic bosonic and
 fermionic wave-functions determine the phase shift, and since the
 asymptotic supercharge takes the bosonic wave-function to the
 fermionic wave-function (and vice versa), we see that the difference
 in spectral densities in these circumstances is determined by the
 asymptotic supercharge. It can be checked in explicit examples that
 this expectation is borne out.

 Next, we study how this elementary phenomenon in supersymmetric quantum mechanics has far reaching
 consequences in  higher dimensional generalizations of the weighted trace.

\subsection{The elliptic genus}
\label{geneg}

Elliptic genera of superconformal field theories in two dimensions can be defined
as twisted toroidal path integrals in the sector with periodic boundary conditions
for the right-movers \cite{Witten:1986bf},
\be
\chi=\mbox{Tr}_R (-)^{\bar{F}} q^{L_0-c/24} \bar{q}^{\bar L_0-c/24} x^{Q}~,
\ee
where $q=e^{2\pi i\tau}$, with $\tau$ being the modular parameter of the torus,
and $L_0$ ($\bar L_0$) is the left (right)-moving scaling dimension.
Here, the right-moving fermion number is
denoted $\bar{F}$, and we have introduced a chemical potential $x=e^{2\pi i \gamma} $
for a $U(1)$ symmetry $Q$ that commutes with the right-moving supercharge.
Since the trace corresponds to a toroidal path integral, it is guaranteed to have good modular
covariance properties, as well as good periodicity properties in the twist variable. To define
the elliptic genus, it is sufficient to have $(0,1)$ supersymmetry in two dimensions.

As far as the right-moving primaries are concerned, the elliptic genus behaves like a weighted
trace in supersymmetric quantum mechanics. As a consequence, when the conformal field theory
has a discrete spectrum of right-moving primaries,  the elliptic genus projects
 onto the right-moving ground states. The result is then a holomorphic
Jacobi form.

The elliptic genus of theories
with a spectrum of conformal dimensions that contains a continuum,
obtains contributions from right-moving ground states, as well as from
the continuum, due to the difference
in the density of bosonic and fermionic right-moving primaries. This difference is
determined in terms of the asymptotic right-moving supercharge \cite{Troost:2010ud,Ashok:2013kk}.
The result is a mock modular
contribution from the right-moving ground states, and a modular completion arising from the continuum \cite{Troost:2010ud}.
These contributions are inextricably linked by modularity and ellipticity \cite{Zwegers,Zagier}.
This mathematical phenomenon has a counterpart in physics, where fixing the asymptotic geometry
can lead to a unique filling in in the interior, and thus to a unique set of right-moving
ground states.

We wish to comment on an attractive feature of the elliptic genus.
 Superconformal field theories with continuous spectra exhibit a
volume divergence in their partition function. When the global symmetry group is
as large as the volume divergence, one can use symmetries to extract the
divergent factors, and remain with a calculable and well-defined finite factor. When this
is not the case, one can use a twisted partition function, the elliptic genus, to
eliminate the volume divergence.
The integral over the difference in spectral densities
can be finite, even in the presence of infinite volume. Thus, the elliptic genus
can serve as a modular covariant infrared regulator.

Next, we review how these generic features are realized in the cigar coset superconformal
field theory.

\subsection{The cigar conformal field theory}
\label{cigarcft}
The cigar geometry \cite{Elitzur:1991cb,Mandal:1991tz,Witten:1991yr,Dijkgraaf:1991ba}
 and dilaton $\Phi$ associated with the $SL(2, \mathbb{R})_k/U(1)$ model take the form
\be\label{cigar}
ds^2=2k \tanh^2 \left(\frac{\rho}{\sqrt{2 k}}\right) d\theta^2
+d\rho^2~,~~~~
\exp(2 \Phi )= \frac{g_{0}^2}{\cosh^2 \left(\frac{\rho}{\sqrt{2 k}}\right)}~.~~~~
\ee
The angular direction $\theta$ has periodicity $2\pi$ to ensure
smoothness of the background at the tip. We have chosen the convention
$\alpha'=2$.

Vertex operators in the $SL(2, \mathbb{R})_k/U(1)$ $N=2$ superconformal field theory are determined by
five quantum numbers ($n_f, \bar{n}_f, m, \bar{m}, j$);
we have
$n_f=F-1 \, (1/2)  ~\mbox{in the NS (R) sectors},$
where $F$ is the left-moving worldsheet fermion number,
and  a similar formula holds for the right-moving fermion numbers;
the quantum numbers $m, \bar{m}$ are related to the momentum $p$ and winding $w$ around the angular
direction $\theta$ of the cigar  via
(see e.g. \cite{Giveon:2003wn} for details)
\be\label{mmpw}
(m+n_f, \bar{m}+\bar{n}_f)=\frac12 (p+kw,-p+kw),~\qquad p\in Z+n_f-\bar n_f~, \,w\in Z~.
\ee
Roughly speaking, the quantum number $j$ is associated with the momentum in the radial direction $\rho$
of the cigar.
More precisely, the possible values of $j$ are inherited from either
the continuous or the discrete representations of $SL(2,\mathbb{R})$.
In the continuous representations,
\be\label{js}
j=-1/2+is~, ~~s\in \mathbb{R}~,
\ee
while in the discrete representations,
\be
j=|m|-n~, ~~~~n=1,2,...~,
\ee
and similarly for the right movers.
Normalizability of the discrete states implies that they are dominated by their
winding contribution, namely,
\be
{\rm sign}(m+n_f)={\rm sign}(\bar m+\bar n_f)~.
\ee
Moreover, in the quantum theory,
we have to impose the unitarity bound \cite{Giveon:1999px,Maldacena:2000hw}
\be
-\frac12 < j < \frac{k-1}{2}~.
\ee
The scaling dimension and R-charge are determined by these quantum numbers,
\be\label{sc}
h=\frac{(m+n_f)^2-j(j+1)}{k}+\frac{n^2_f}{2}~,~~~~~~R=\frac{2(m+n_f)}{k}+n_f~,
\ee
with similar equations for the right movers.
Since the $SL(2, \mathbb{R})_k/U(1)$ conformal field theory contains both discrete and continuous states,
the features described in the previous subsections are relevant, and have
 interesting implications.
\subsection{The cigar elliptic genus}
In the cigar superconformal field theory, we have additional symmetries compared
to the generic case discussed in subsection \ref{geneg}. We have a left-moving superconformal
algebra with left-moving R-charge $R$ as well as a global $U(1)$ symmetry $P$ corresponding to angular
momentum in the direction $\theta$. Since both commute with the right-moving supercharge $\bar{G}_0$, they can
both be used to twist the elliptic genus.
We study the trace with periodic boundary conditions for both left- and right-movers and with the
twists
\be
\chi=\mbox{Tr}_R (-)^{F+\bar{F}} q^{L_0-c/24} \bar{q}^{\bar L_0-c/24} z^{R} y^P~.
\ee
The operator $F$ denotes left-moving fermion number,
while we have fugacities $z=e^{2\pi i \alpha} $ and $y=e^{2\pi i \beta}$,
where  $\alpha$ and $\beta$ are the chemical potentials for the $U(1)$ R-charge and
$\theta$-angular momentum $P$, respectively.
Far on the cylinder, $\rho\to\infty$,
a good approximation for $P$ is the  linear momentum,
while near the tip, $\rho\to 0$, $P$ is a true angular momentum.

As described in subsection \ref{Wittenind}, the contribution of the
continuum is determined in terms of an asymptotic right-moving
supercharge, and does not depend on many of the detailed properties of
the model. Let's see how this simplification works in practice.
For the $SL(2, \mathbb{R})_k/U(1)$ model, the
reflection amplitudes are given by~\cite{FZZ,Teschner:1999ug,Giveon:1999px}
\be\label{6} R(m,\bar
m,j)=\left(\frac{1}{\pi}\frac{\Gamma(1+\frac{1}{k})}{\Gamma(1-\frac{1}{k})}\right)^{2j+1}
\frac{\Gamma(1-\frac{2j+1}{k})\Gamma(j+1+m)\Gamma(j+1-\bar
  m)\Gamma(-2j-1)} {\Gamma(1+\frac{2j+1}{k})\Gamma(m-j)\Gamma(-j-\bar
  m)\Gamma(2j+1)}~, \ee
 and since $R_{bos}/R_{ferm}\equiv R(\bar
n_f=-1/2)/R({\bar n_f=1/2})$, with $p,w$ and $n_f$ held fixed in equation
(\ref{mmpw}), one finds for the measure on the real line
\cite{Ashok:2011cy}
\be \Delta\rho=-\frac{1}{\pi} \frac{1}{2is-p+kw}~.
\ee
Indeed, asymptotically on the cylinder we have the right-moving supercharge
$\bar{G}_0\sim (2is-p+kw)$ in the Ramond sector.\footnote{To be more precise,
asymptotically, the right-moving supercurrent can be written as: $\tilde{G}^+
= \frac{i}{2} (\tilde{\psi}_\rho + i \tilde{\psi}_\theta) \partial(\rho - i \theta)
+ \frac{i}{\sqrt{2k}} \partial (\tilde{\psi}_\rho + i \tilde{\psi}_\theta)$. For large level $k$, the last
term drops out. The zero mode sector then gives rise to the right-moving supercharge
that we quote, after stripping the fermion zero mode.}
Thus, the spectral asymmetry $\Delta\rho$ is fully determined by the
asymptotic right-moving supersymmetry charge
\cite{Akhoury:1984pt,Troost:2010ud,Ashok:2013kk}. This property will play a key role
in the discussion about the universality of our results.

Let us note further that the unique
filling in of the internal geometry based on fixed asymptotics,
mentioned in subsection \ref{geneg}, is discussed in detail for the cigar case in
\cite{Hori:2001ax}.\footnote{Other examples to which similar methods should apply include
the asymptotic linear dilaton spaces of \cite{Kiritsis:1993pb}--\cite{Murthy:2013mya}.}

\section{The large level limit of the elliptic genus}
\label{largelevel}
In this section, we analyze the large level limit of the elliptic genus of the cigar
conformal field theory. We identify the states that contribute to it, their wave-functions
and where they are localized.
\subsection{The limit in the Hamiltonian form}
We will calculate the large level $k$ limit of the \eg of
the $SL(2, \mathbb{R})_k/U(1)$ coset model in two ways.  The
starting point for both is the functional integral
expression for the \eg at finite level $k$ that was calculated in
\cite{Troost:2010ud, Eguchi:2010cb, Ashok:2011cy}:
\be
\label{re}
 \chi = k\int_{0}^{1} ds_1 ds_2 \sum_{m,w\in \mathbb{Z}}
\frac{\theta_{11}(s_1 \tau +s_2-\alpha
  (k+1)/k+\beta,\tau)}{\theta_{11}(s_1 \tau +s_2-\alpha
  /k+\beta,\tau)} e^{2\pi i\alpha w}
e^{-\frac{k\pi}{\tau_2}|(m+s_2)+(w+s_1)\tau|^2}.  \ee
This result can be understood as arising from a $U(1)$ gauged
linear sigma model with one charged $N=(2,2)$  chiral superfield giving rise to the
ratio of theta functions, as well as a St\"uckelberg superfield that cancels the
vector multiplet contribution except for their holonomies $s_{1,2}$, and that
contributes the last, zero-mode factor depending on windings $m$ and $w$ \cite{Ashok:2013zka,Ashok:2013pya,Murthy:2013mya}.
We set the chemical potential for angular momentum $\beta=0$
in this section. We shall turn it on occasionally to make specific points.  Since equation
(\ref{re})
arises from path integration over a toroidal worldsheet \cite{Troost:2010ud}, the
functional integral expression is by construction a Jacobi form; that is
\be\label{tra} \chi(\tau+1,\alpha) =
\chi(\tau,\alpha),~~~~\mbox{and}~~~~ \chi(-1/\tau,\alpha/\tau) =
e^{\pi i \hat{c} \alpha^2/\tau}\chi(\tau,\alpha), \ee with \be
\hat{c}\equiv c/3=1+\frac{2}{k}~.  \ee

A first way to find the large level genus, is to start from the Hamiltonian
form of the elliptic genus $\chi$ that is obtained from (\ref{re}) after
manipulations that lay bare the contributions as arising from
a physical Hilbert space \cite{Troost:2010ud, Eguchi:2010cb, Ashok:2011cy}.
The second
way is to take the large $k$ limit directly at the functional integral
level (\ref{re}). We follow the latter route in section \ref{saddle}.  Each derivation
will teach us various physical lessons.

After double Poisson resummation and performing the integrals over the holonomies
$s_1$ and $s_2$, the genus
(\ref{re}) can be written in the following way
\be
\chi = \chi^{hol} + \chi^{rem}~,
\ee
where $\chi^{hol}$ is the holomorphic contribution from right-moving ground states
 and $\chi^{rem}$ contains non-holomorphic pieces arising from the continuum. They read~\footnote{We consider
here the cases with integer level, $k=1,2,...$.}
\be
\label{holopart}
\chi^{hol} = \frac{i \theta_{11}(z,q)}{\eta^3}
\sum_{\gamma=0}^{k-1} \sum_{m \in \mathbb{Z}}
\frac{q^{k m^2- m \gamma} z^{2 m - \frac{\gamma}{k}}}{1-z q^{km -\gamma}}~,
\ee
and
\be
\chi^{rem} =
- \frac{i \theta_{11} (z,q)}{\pi \eta^3} \sum_{p,w}
\int_{- \infty - i \epsilon}^{+ \infty - i \epsilon}
 \frac{ds}{2is-p+kw} q^{ \frac{s^2}{k} + \frac{(kw+p)^2}{4k}} z^{ \frac{kw+p}{k}}
\bar{q}^{ \frac{s^2}{k} + \frac{(kw-p)^2}{4k}}~.
\ee
Taking the large $k$ limit of these expressions, one obtains
\be\label{large}\chi(k=\infty)=
{\frac{i\theta_{11}}{\eta^3}}\left[{\frac{1}{2}}{\frac{1+z}{1-z}}+\sum_{n=1}^\infty
\left({\frac{zq^n}{1-zq^n}}-(z\to z^{-1})\right)+\frac{i\alpha}{2 \tau_2} \right].
\ee
Note that this expression is finite at infinite level $k$. Namely, in
the large $k$ expansion of the elliptic genus, there are no
terms that scale like a positive power of the level $k$.  This is
an indication that this expression is universal.
 The non-holomorphic piece,
$\frac{i \theta_{11}}{ \eta^3}\frac{i\alpha}{2 \tau_2}$, is due to the
continuous representations, as discussed in subsection \ref{geneg}.
Both the holomorphic and the non-holomorphic
contributions are essential for modular covariance. Though this follows
from a similar statement at any finite level \cite{Troost:2010ud, Eguchi:2010cb, Ashok:2011cy},
it is also instructive to show this
directly on the infinite level expressions.

To demonstrate that the limit elliptic genus
(\ref{large}) transforms according to the rules
(\ref{tra}) with $\hat{c}=1$, we first observe that the holomorphic part of the genus (\ref{large})
can be written in the following form (see useful formulas in the appendix):
\be
{\frac{i\theta_{11}}{\eta^3}}\left[{\frac{1}{2}}{\frac{1+z}{1-z}}+\sum_{n=1}^\infty
\left({\frac{zq^n}{1-zq^n}}-(z\to z^{-1})\right)\right]= - \frac{1}{2 \pi} \frac{\partial_{\alpha}\theta_{11}}{ \eta^3}~.
\ee
Consider the modular S-transformation of the holomorphic contribution:
\begin{eqnarray}
\label{anomaltransfo}
- \frac{1}{2 \pi} \frac{\partial_\alpha \theta_{11}(z,q)}{\eta^3}
& \rightarrow &
- \frac{1}{2 \pi} e^{ \pi i \frac{\alpha^2}{\tau}} \frac{ \partial_\alpha \theta_{11}(\alpha,\tau)}{\eta^3}
- \frac{\alpha}{\tau}
e^{ \pi i \frac{\alpha^2}{\tau}} \frac{i \theta_{11}(\alpha,\tau)}{\eta^3} \, .
\end{eqnarray}
The anomalous transformation of the non-holomorphic term in equation (\ref{large}),
\begin{eqnarray}
\frac{i \theta_{11}}{
  \eta^3}\frac{i\alpha}{2 \tau_2} & \rightarrow &
e^{ \pi i \frac{\alpha^2}{\tau}} \frac{i \theta_{11}}{
  \eta^3}\frac{i\alpha}{2 \tau_2} \frac{\bar{\tau}}{\tau} =
e^{ \pi i \frac{\alpha^2}{\tau}} \frac{i \theta_{11}}{
  \eta^3}\frac{i\alpha}{2 \tau_2}
+e^{ \pi i \frac{\alpha^2}{\tau}} \frac{i \theta_{11}}{
  \eta^3} \frac{\alpha}{\tau} \, ,
\end{eqnarray}
exactly cancels the second term in
the right hand side of transformation rule
(\ref{anomaltransfo}), and renders the genus (\ref{large}) modular covariant.
Indeed, the derivative we take can be viewed as a modular covariant derivative acting
on a Jacobi form. This is an elementary
incarnation of how shadows and mock Jacobi forms are bijectively linked in favorable
circumstances \cite{Zwegers,Zagier,Dabholkar:2012nd}.

\subsection{The identification of states}
\label{identification}
We saw that the holomorphic piece of the elliptic genus of the cigar at $k\to\infty$
takes the form
\be\label{aaaa}
\chi^{hol} = {\frac{i\theta_{11}(z,q)}{\eta^3}}\left[{\frac{1}{2}}{\frac{1+z}{1-z}}+\sum_{n=1}^\infty
\left({\frac{zq^n y^{n}}{ 1-zq^n}}-(z\to z^{-1},y\to y^{-1})\right)\right]~.\ee
Here we have added in the dependence on the chemical potential for momentum along the angular direction
$\theta$.
This gives extra information helpful
in identifying the states that contribute.

The various terms in equation (\ref{aaaa})
amount to the modules of states that are right-moving ground
states in the Ramond sector,
and which survive the $k\to\infty$ limit.
Concretely, in the notation of subsection \ref{cigarcft},
these are discrete states with 
quantum numbers $(n_f,\bar n_f,m,\bar m,j)$, satisfying $|\bar m|=j+1$.
In terms of the asymptotic winding $w$ on the cigar's throat,
the only right-moving ground states which do
not decouple in the large level limit are winding plus or minus one states.

States with $w=1$, $p=1,2,...$ and $\bar n_f=-1/2$ give rise to the
${\frac{i\theta_{11}}{\eta^3}}\sum_{n=1}^\infty {\frac{z y^nq^n}{1-zq^n}}$ term in the elliptic genus,
where the integer $p=n=1,2,...$ in the sum is the momentum
on the cigar, for the asymptotic winding one states,
 and the theta-function factor,
$i\theta_{11}=\sum_{n_f\in
\mathbb{Z}+1/2}(-1)^{n_f+1/2}q^{n_f^2/2}z^{n_f},$
is a sum over the left-moving fermion number, $n_f$.
The conjugates of these right-moving ground states, those with
winding
$w=-1$ and momentum $p=-n=-1,-2,...$, as well as opposite right-moving fermion number, $\bar n_f=1/2$,
contribute to the term in the elliptic genus obtained by mapping $z\to z^{-1}$ and $y\to y^{-1}$.

We can also interpret these contributions as arising
from  $N=2$ superconformal characters at central charge $c=3$ and with
quantum numbers $Q = \pm 1 + n_f$ and $h=\pm p+c/24$,
along the lines described in~\cite{Giveon:2013ica}.

The Hamiltonian interpretation of the term
${\frac{i\theta_{11}}{\eta^3}}{\frac{1}{2}}{\frac{1+z}{1-z}}$ in the elliptic genus is more subtle, since
as it stands it contains half-integer multiplicities. The subtlety arises from the ambiguity in interpreting
a state at the bottom of the continuum as a right-moving ground state. We rewrite this term
as
\begin{eqnarray}\label{di}
{\frac{i\theta_{11}}{\eta^3}}{ \frac{1}{2}}{\frac{1+z}{1-z}}
&=& {\frac{i\theta_{11}}{\eta^3}} \frac{z}{1-z} + \frac{1}{2}  {\frac{i\theta_{11}}{\eta^3}} \, .
\end{eqnarray}
We can interpret the first contribution as arising from a primary with
$w=1$ and zero angular momentum $p=0$. It is a marginal winding
tachyon -- the $N=2$ Liouville operator, combined via FZZ duality with
the dilaton deformation (see e.g.~\cite{Aharony:2004xn} and references therein for details).
The second contribution is the result of
integration over the continuum of states, and in particular is
associated to a choice of regularization of this integral at zero
radial momentum $s=0$. To arrive at our Hamiltonian interpretation of
the last term, we broke $\alpha \leftrightarrow - \alpha$ charge
conjugation symmetry. Another, charge conjugate Hamiltonian
interpretation is equally possible.

\subsection{Target-space location}
\label{location}

We focus now on the target-space interpretation of the primary states.
In particular, we determine whether these states are located near the  string scale
tip of the cigar, $\rho \lesssim 1$, or at the cap of the cigar
and the regime where it is glued to the asymptotic cylindrical throat,
$\rho \lesssim \sqrt{k}$ (see figure \ref{tipcapglue}).
There are two ways to tackle this important issue. The first is direct,
by computing wave-functions corresponding to the states. The second is indirect,
by analyzing the localization of the state contributions in the holonomy plane, as
we will do in section \ref{uvir}.

In the previous section, we found the $SL(2, \mathbb{R})_k/U(1)$ coset
quantum numbers associated with the primary states that contribute to the
holomorphic term in the elliptic genus.  With this information at hand, we
can translate the conformal dimension into a Laplacian eigenvalue
for the Laplacian on the exact supersymmetric cigar background
 (\ref{cigar}) (see e.g.  \cite{Dijkgraaf:1991ba}
for the application of this technique to the bosonic black hole). We find the
target-space wave-function of our states using their conformal dimension~\footnote{Here
we work in the NS-NS sector with $n_f=\bar n_f=0$.}
\be\label{vc}
L_0 |\Psi \rangle = h |\Psi \rangle ~~~~\mbox{and} ~~~~\bar{L}_0 |\Psi \rangle = \bar{h} |\Psi \rangle,
\ee
with scaling operators
\be
L_0=-\frac{1}{k} (\triangle -m^2),~~~~\bar{L}_0=-\frac{1}{k} (\triangle -\bar{m}^2),
\ee
where the $SL(2, \mathbb{R})$ Laplacian $\triangle$ takes the
form\footnote{Alternatively,
we could directly use the Laplacian on the cigar geometry. Here, we take a coset
conformal field theory point of view, i.e. we take the parent conformal field theory
as our starting point.}
\be
\label{laplace}
\triangle =\frac{k}{2} \partial_{\rho}^2+\sqrt{k/2}\coth(\frac{\rho}{\sqrt{k/2}}) \partial_{\rho}-\frac{1}{\sinh^2(\frac{\rho}{\sqrt{k/2}})}\left( m^2 +\bar{m}^2 -2 \cosh(\frac{\rho}{\sqrt{k/2}})m \bar{m} \right).
\ee
We concentrate on the winding numbers $|w|=1$ corresponding to the BPS primaries that survive
the infinite level limit.
Taking the large $k$ limit at fixed $\rho$ of the condition
\be\label{aaaaaa}
(L_0 +\bar{L}_0) |\Psi \rangle= (h+\bar{h}) |\Psi \rangle~,
\ee
we find that the BPS primary states with $|w|=1$ satisfy the following equation
\be\label{tv}
\left( -\partial_{\rho}^2 - \frac{1}{\rho} \partial_{\rho} +
\frac{\rho^2}{4} -1 +\frac{p^2}{\rho^2}\right) |\Psi_p \rangle = |p| |\Psi_p \rangle~.
\ee
This equation describes a two-dimensional harmonic oscillator with frequency
$1$ and angular momentum $L=p$ (see~\cite{Giveon:2013ica} for more detail).
The difference $L_0 -\bar{L}_0$ is fixed to $h-\bar{h}=p w =|p|$.
%
In terms of the energy, $E$, and angular
momentum, $L$, we see that states that contribute to the \eg
(\ref{aaaa}) have
\be E=|L|=n=0,1,2,...~.  \ee
Since the level $k$ dropped out of equation (\ref{tv}), it follows
that these states are localized at the tip of the cigar.

We conclude that modularity
of the \eg implies the existence of these discrete states which in turn makes
the origin of $\mathbb{R}^2$, as obtained by taking the infinite level limit of the cigar,
special.  In ordinary flat space $\mathbb{R}^2$, all states with
definite left and right conformal dimension $h$ and $\bar{h}$ are smeared all over the
two-plane.  Equation (\ref{tv}),
governing our wave-functions,
implies that at the tip of the cigar there are
states with fixed left- and right-moving conformal dimensions that are localized at
the string scale.

\section{ Universality}
\label{universality}

As explained in section \ref{generalities}, the non-holomorphic term in the \eg is due
to a mismatch, $\Delta \rho$, between the densities  of worldsheet
bosonic and fermionic degrees of freedom in the continuum.
In the limit $k\to\infty$, locally the difference in densities
approaches zero, $\Delta\rho\sim1/k$, but
since the area of the cigar's cap blows up, $V_{cap}\sim k$,
there is a finite contribution that scales like $\alpha/\tau_2$.
Hence, from the target-space point of view the non-holomorphic piece is an infrared effect.
Moreover, as emphasized in section \ref{generalities},
the spectral asymmetry $\Delta\rho$ is completely determined
by supersymmetry and the asymptotic geometry;
that is by the topology of the cigar and not by
the details of how its geometry is filled in.
Hence, it is natural to expect this large $k$ non-holomorphic piece  to be universal.

The topology of the cigar also dictates that the Witten index, $\chi(\alpha=0)$,
should be equal to $1$.
As implied from the discussion around  equation (\ref{di}),
this too is due to a supergravity mode -- the multiplet of the zero mode of the dilaton.
Thus, as expected, all aspects of the \eg that are determined solely by the topology are controlled
by infrared physics in the target space.

Modularity and the non-holomorphic term determine  the holomorphic term.
As discussed  in the previous sections,
this term is due to discrete states that are located at the tip of the cigar.
Namely, from the target-space point of view this contribution is a UV effect.
Thus, the \eg is another example where a modular transformation
mixes between UV and  IR degrees of freedom in the target space.

Since the non-holomorphic term is universal and since it determines the holomorphic piece via modularity,
we conclude that the holomorphic piece is universal as well. In particular,
we expect to find discrete states at the tip associated with generic large mass black holes.


The way the holomorphic term was obtained in the previous section may look non-universal;
it is a particular sum of
characters of the $N=2$ superconformal algebra.  This sum
is restricted at large level $k$ to asymptotic winding (plus or minus) one
states and includes the denominators that are induced
by their null state structure.

In the rest of this section, we review an alternative route
to modeling the elliptic genus of the cigar theory \cite{Ashok:2013kk},
in terms of a supersymmetric quantum mechanics
that arises after Scherk-Schwarz reduction \cite{Scherk:1978ta}
of the $(2,2)$ supersymmetric sigma model on the angular direction.
It may shed light on the universal nature of the result. In
 \cite{Ashok:2013kk}, two counting functions were matched. On the one hand, the
elliptic genus after factoring out a generic oscillator contribution
$i\theta_{11}/\eta^3$ corresponding to {\em all} generators of the $N=2$
superconformal algebra, and on the other hand, the number of
right-moving ground states as coded in the supersymmetric quantum
mechanics corresponding to given winding $w'$, and with momentum $p$.

Technically, the right-moving ground states contribution to the
elliptic genus was found to be of the form
\begin{eqnarray}
\chi^{hol} &=&  \frac{i \theta_{11}}{\eta^3} 
\left( \sum_{p \ge 0, -p+ kw' \ge 0}- \sum_{p \le -1, -p+kw' \le -1}\right)
z^{\frac{p+kw'}{k}} q^{pw'} y^p \, ,
\end{eqnarray}
which matches the expression (\ref{holopart}) for  the elliptic genus under the identifications
\begin{eqnarray}
p &=& km - \gamma~,
\nonumber \\
w' &=& m + r~,
\end{eqnarray}
where $r$ is a quantum number arising from expanding the null vector denominator in our $N=2$
superconformal characters
in the regime $|q|<|z|<1$, and $m$ and $\gamma$ are the quantum numbers appearing in
equation (\ref{holopart}).
In the large level limit, and ignoring ambiguous contributions at zero winding or zero momentum, we then find the
expression
\begin{eqnarray}
\chi^{hol} &=&  \frac{i \theta_{11}}{\eta^3} \left( \sum_{p \ge 1, w' > 0}- \sum_{p \le -1, w' \le -1}\right)
z^{w'} q^{pw'} y^p
\nonumber \\
&=&  \frac{i \theta_{11}}{\eta^3} \left( \sum_{p \ge 1 }
\frac{z q^p y^p}{1-zq^p}- \sum_{p \ge 1} \frac{z^{-1} q^p y^{-p}}{1-z^{-1} q^p}\right)
\, .
\end{eqnarray}
We conclude that the states with winding $w=+1$ on the cigar are
captured by the states with positive momentum $p \ge 1$ and {\em any}
positive winding $w' \ge 1$ in the supersymmetric quantum mechanics
(and a similar statement holds for the charge conjugate states). In
the supersymmetric quantum mechanics, we ignore the existence of the
fermionic null vector of the $N=2$ superconformal algebra, effectively
adding back in a fermionic degree of freedom. In the supersymmetric
partition function, this is compensated by adding in a bosonic degree
of freedom as well, that gives rise to the infinite sum over winding
$w'$.

This embedding of the calculation of the elliptic genus
directly in flat space may provide an important
hint towards finding a universal $c=3,~N=2$ conformal field theory that describes string theory
at the tip of the cigar.


We also wish to revisit the issue of the size and location in the
target space in the framework of the supersymmetric quantum
mechanics. In \cite{Ashok:2013kk}, target-space wave-functions
for all these states  in the NS-R sector were found by solving the
right-moving ground state Dirac equation $\bar{G}_0|\Psi \rangle =0$ at any finite level. We can take the large
level limit on those wave-functions to find that they localize
on a disk of string scale size in agreement with the above discussion.

Alternatively, we can work in the NS-NS incarnation of the quantum mechanics problem by solving the
Klein-Gordon equation
\be\label{tvtwo}
\left( -\partial_{\rho}^2 - \frac{1}{\rho} \partial_{\rho} +
\frac{{w'}^2}{4} \rho^2
-1 +\frac{p^2}{\rho^2}\right) |\Psi_{w',p} \rangle =pw' |\Psi_{w',p} \rangle  \, ,
\ee
to reach the same conclusion. Here, $L_0-\bar{L}_0 = pw'$, and we have used the Laplace
operator  (\ref{laplace}) for a BPS primary state with generic winding number $w'$.

\subsubsection*{Universality and Euclidean Schwarzschild}

A particularly interesting case to which to apply the concept of universality
 is that of a  large mass $M$ limit of the
Euclidean Schwarzschild black hole. When concentrating on the degrees of freedom that do not
depend on the two-sphere section (namely, when concentrating on the s-wave, and
neglecting string oscillations on the two-sphere), we find again a cigar geometry. We may then expect to find
a similar non-holomorphic piece, arising from the gluing of the cap regime
to the asymptotic cylindrical throat shape
\begin{figure}
\centering
\vspace{-30mm}
\includegraphics[width=0.7 \textwidth]{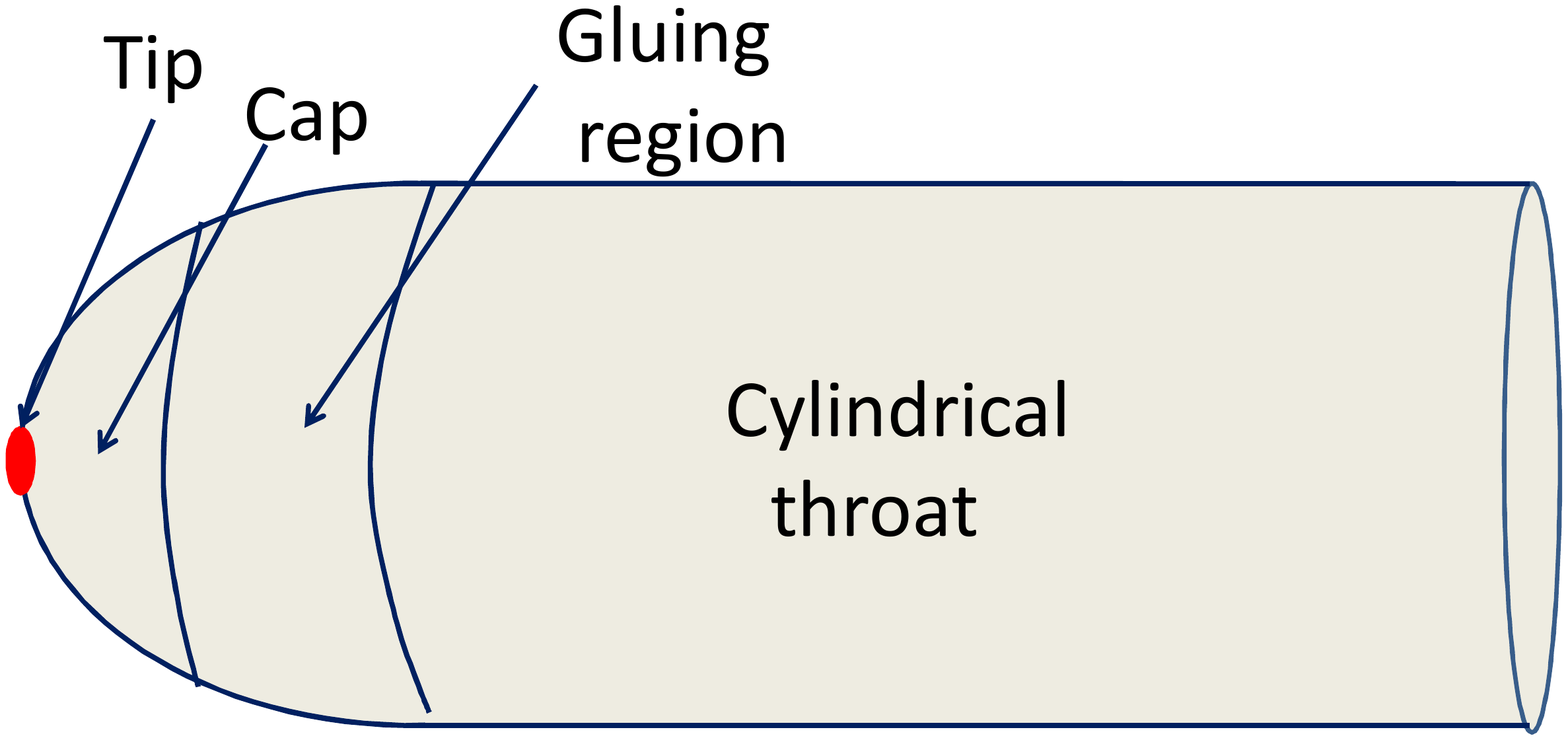}
\caption{In stringy units,
the size of the tip is order 1,
and the size of the cap of the $SL(2,\mathbb{R})_k/U(1)$ (mass $M$ Schwarzschild)
cigar is of order $\sqrt{k}$ ($M$).
A rough definition of the cap is the region at which the curvature (in the string frame)
is of order $1/\sqrt{k}$ ($1/M$).
Deep in the semi-infinite cylindrical throat the curvature vanishes.
The drop in curvature from $1/\sqrt{k}$ ($1/M$)
to zero happens exponentially fast in the gluing region.
The universality discussed in this section refers to the cap,
which in the large $k$ ($M$) limit looks like $\mathbb{R}^2$
with the addition of states that live at the tip
and a small mismatch in the continuous spectrum, $\Delta\rho$.}
\label{tipcapglue}
\end{figure}
(see figure \ref{tipcapglue}). It would be dictated
by the right-moving supercharge at infinity on the Schwarzschild geometry.
Once we fixed the non-holomorphic piece, worldsheet modularity
implies that the holomorphic piece is universal as well.
Namely, the set of right-moving ground
states will be fixed, and may contain a spectrum of discrete states localized near the tip,
that is near the Lorentzian horizon.

More speculatively still,
we may define an analog of the elliptic genus for a $(0,1)$
two-dimensional conformal field theory on the Euclidean Schwarzschild
background at finite mass $M$. The Ricci flat geometry is reliable at leading order
in $1/M$. For this model, there is a twisted trace of the type reviewed
in subsection \ref{geneg} \cite{Witten:1986bf}. It would be interesting to compute the
contribution to such a twisted trace from the continuum and the degrees
of freedom associated to the asymptotic geometry, including the
two-sphere, at finite mass. By combining modularity
(i.e. universality of the modular completion), as well as a unique filling
in of the asymptotics with given topology and satisfying Einstein's
equations, one may be able to draw far reaching conclusions on degrees
of freedom localized in the deep interior, near the tip. It would be
interesting to execute this program in detail.

\section{The saddle point approach}
\label{saddle}
In the previous section, we analyzed the large level limit of the
cigar genus in a Hamiltonian approach.  In this section, we will take
the limit in a Lagrangian formalism, through saddle point analysis.
This approach sheds light, amongst other things, on the fact that the
holomorphic piece of the \eg can be written
as a derivative in the large level limit.

\subsection{The holonomy plane}
\label{planarize}
Periodicity properties of theta functions (see the appendix),
allow us to extend the range of integration over the holonomies $s_1$ and $s_2$ in
equation (\ref{re}) to obtain
\begin{eqnarray}
\chi=  k \int_{-\infty}^{+\infty}
ds_1 ds_2  \frac{ \theta_{11} (s_1 \tau + s_2 - \alpha - \alpha/k ,\tau)}{
\theta_{11} ( s_1 \tau + s_2 - \alpha/k ,\tau)}
e^{ - \frac{k \pi}{\tau_2} | s_2+ s_1 \tau|^2 } \, .
\end{eqnarray}
In the large $k$ limit, we can do a saddle point approximation
to the integral. We must be careful in evaluating the potential large
$k$ dependence of the $\theta_{11}$ functions. To that end,
we redefine the integration variables,
\be
s_1 \tau + s_2 = x_1+ ix_2~,
\ee
and pick up the measure factor $\tau_2$,
\begin{eqnarray}
\chi
 & = & \frac{k}{\tau_2} \int_{-\infty}^{+\infty}
d^2x  \frac{ \theta_{11} (x_1+ix_2 - \alpha - \alpha/k ,\tau)}{
\theta_{11} ( x_1+ix_2 - \alpha/k ,\tau)}
e^{ - \frac{k \pi}{\tau_2} (x_1^2+x_2^2) } \, .
\end{eqnarray}
Rescaling  the holonomy plane variables,
we get
\begin{eqnarray}
\chi
 & = &  \int_{-\infty}^{+\infty}
d^2x \frac{ \theta_{11} (  \sqrt{\frac{\tau_2}{k}} (x_1+ix_2) - \alpha - \alpha/k ,\tau)}{
\theta_{11} ( \sqrt{\frac{\tau_2}{k}} (x_1+ix_2)  - \alpha/k ,\tau)}
e^{ - \pi (x_1^2+x_2^2) } \, .
\end{eqnarray}
This expression is a good  starting point of the saddle point analysis we perform below.
\subsection{The saddle point analysis}
In the large $k$ limit, we can expand the $\theta$
functions in the denominator and numerator to get 
\begin{eqnarray}\label{12}
\chi
 & = &  \int
d^2x  \frac{ \theta_{11} ( \alpha,\tau) - (\sqrt{\frac{\tau_2}{k}} (x_1+ix_2) - \alpha/k )
\partial_\alpha \theta_{11}(\alpha,\tau) + \dots  }{
2 \pi (\sqrt{\frac{\tau_2}{k}} (x_1+ix_2) - \alpha/k + \dots) \eta^3   }
e^{ - \pi (x_1^2+x_2^2) } \, .
\end{eqnarray}
The validity of the expansion implies that the range of integration is
$x_1^2+x_2^2<\frac{k}{\tau_2}$.  In the large $k$ limit, however, the
exponential suppression renders this bound irrelevant and the range of
integration is the whole $(x_1,x_2)$ plane. The leading term
scales like $\sqrt{k}$,
\be
\chi_{lead} = 
\sqrt{\frac{k}{\tau_2}}\frac{\theta_{11} ( \alpha,\tau)}{\eta^3   }\int \frac{d^2x}{ x_1+ix_2}
e^{ - \pi (x_1^2+x_2^2) } \, .
\ee
This term, however, vanishes because of the  integration over the angular direction.
Namely, writing $x_1+ix_2=r e^{i\theta}$, we see that the integral over the angle $\theta$ vanishes.

At the next order we find two terms. The simplest one is from the second term in the numerator,
\begin{eqnarray}
\chi_{1} &=& -\frac{ \partial_\alpha \theta_{11} (\alpha,\tau) }{ 2 \pi \eta^3} \, \int d^2x  e^{ - \pi (x_1^2+x_2^2) }=-\frac{ \partial_\alpha \theta_{11} (\alpha,\tau) }{ 2 \pi \eta^3}~.
\end{eqnarray}
This term is nothing but the holomorphic part of the genus
(\ref{large}). Note that the contribution to this term comes from a
region of radius of order $1$ in the $(x_1,x_2)$ plane.

The second  contribution comes from the first term in the numerator and a little disk of size
$\alpha/\sqrt{k \tau_2}$ around the origin.
For this contribution, we have
\begin{eqnarray}
r< \alpha/\sqrt{k \tau_2}~.
\end{eqnarray}
 That integral, after expanding the denominator
{\em in this region}, boils down to
\begin{eqnarray}
-2 \pi \int_0^{\frac{\alpha}{\sqrt{k \tau_2}}} dr\, r
\frac{k}{\alpha} e^{- \pi r^2} &=& \frac{k}{\alpha} (-1+ e^{- \frac{\pi \alpha^2}{k \tau_2}})
\nonumber \\
&\to& - \frac{\alpha \pi}{\tau_2}~,
\end{eqnarray}
such that we get a contribution
\begin{eqnarray}
\chi_{non-hol} &=& \frac{\theta_{11}}{\eta^3} (- \frac{\alpha}{2 \tau_2}) \, ,
\end{eqnarray}
which is the non-holomorphic term in the genus
(\ref{large}). It can be checked that the other region of integration
(for radii larger than the critical radius $\alpha/\sqrt{k \tau_2}$)
does not contribute due to the angular integration.

The regions in the
holonomy plane that contribute to the two terms in the elliptic
genus demand a physical interpretation,
perhaps in terms of a model living on the holonomy plane.
We will come back to this topic shortly.

We note that the saddle point approach can be used as a technique to tackle
many questions in  the $SL(2, \mathbb{R})_k/U(1)$ conformal field theory directly at large level.
This includes the calculation of correlation functions,
and also the partition function that we turn to next.

\subsection{The partition function}

In this subsection, we use the saddle point approach to analyze the
partition function at large level. Our main goal here is to extract more
information about the discrete states in the large $k$ limit.
(The \eg is sensitive only to those discrete states that are right-moving
ground states).

There is another reason the partition function is useful
 in the study of the large level limit of the
$SL(2,\mathbb{R})_k/U(1)$ model; it is modular invariant.
This entails that expanding
the partition function in $1/k$ yields modular invariant
expressions at each order.

The functional integral representation of the partition function, including
extra fermionic degrees of freedom of a ten-dimensional superstring theory  is \cite{Sugawara:2012ag} \footnote{For concreteness,
 we consider here e.g. the near horizon Euclidean geometry associated with $k$ near extremal NS5-branes
 in the type II superstring; see e.g. \cite{Sugawara:2012ag,Giveon:2013ica} for details.}
\ben\label{axx}
Z_{\sigma_L,\sigma_R}&=&k  \sum_{m_i\in \mathbb{Z}} \int_0^1 ds_1 ds_2 \epsilon( \sigma_L;m_1,m_2) \epsilon( \sigma_R;m_1,m_2) \times \\ \nonumber & &f_{\sigma_L}(s_1\tau+s_2,\tau)f_{\sigma_R}^{*}(s_1\tau+s_2,\tau)
e^{-\frac{\pi k}{\tau_2} |(s_1+m_1) \tau+(s_2+m_2)|^2}~.
\een
The left- and right-handed spin structures, $\sigma_L, \sigma_R$, run over the four sectors $NS, \tilde{NS}, R$ and $\tilde{R}$. The functions $f_{\sigma}$ are given in terms of standard theta functions via
\be
f_{\sigma}(u,\tau)=\frac{\theta_{\sigma}(u,\tau)}{\theta_1(u,\tau)} \left( \frac{\theta_{\sigma}(0,\tau)}{\eta(\tau)} \right)^3,
\ee
where $\theta_\sigma\equiv\theta_{1,2,3,4}$ for $\sigma=\tilde R,R,NS,\tilde{NS}$, respectively,
and
\be
\epsilon= \left\{\begin{array}{ll}
   1 &,~~~\sigma = NS,\\
   (-)^{m_1+1}&,~~~   \sigma = \tilde{NS},\\
   (-)^{m_2+1} &,~~~\sigma = R, \\
    (-)^{m_1+m_2}  &,~~~ \sigma = \tilde{R} . \end{array}\right.
   \ee
   Formally, this expression is modular invariant. However, it is
   divergent at $s_{1,2}=0 , 1$. The target-space origin of this
   divergence is the infinite volume of the cylinder \cite{Hanany:2002ev}.  The infinite
   volume is an obstruction to cleanly dissecting various
   contributions to the partition function (see e.g. \cite{Israel:2004ir} for related puzzles)
 in contrast to the elliptic
   genus analysis. In the elliptic genus,
     the volume divergence is canceled by a (twisted) right-moving
     fermionic zero mode \cite{Troost:2010ud}. Operationally, a $\theta_{1}$ function in
     the denominator is canceled by an identical $\theta_1$ function
     in the numerator. What remains is a finite
     integral.

\begin{figure}
\centering
\includegraphics[width=0.5\textwidth]{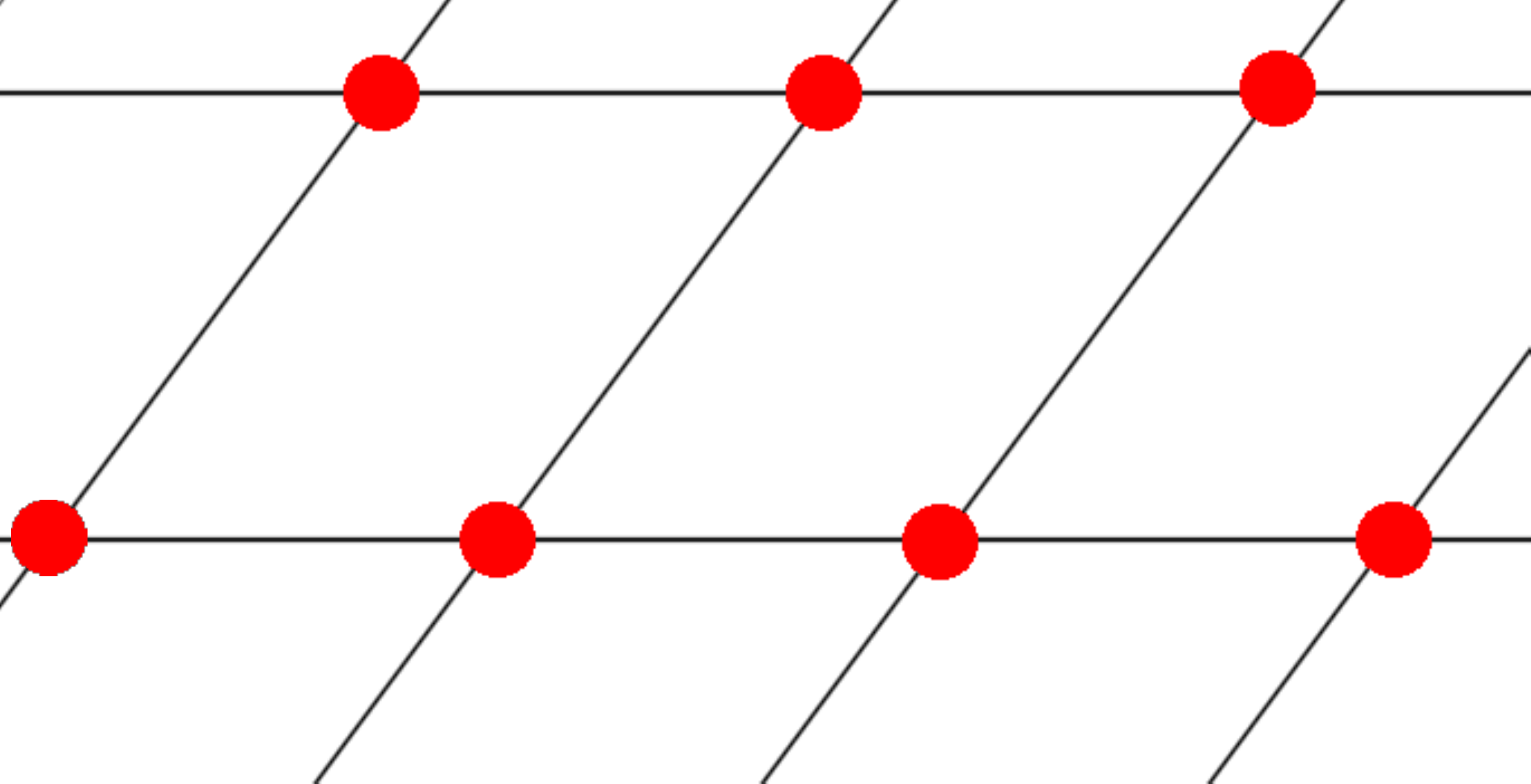}
\caption{We cut out little disks around lattice points in the holonomy plane to regularize
modular invariantly. }\label{cutdisk}
\end{figure}

Below, we follow a different way to regularize the partition function,
which is modular invariant.
Using similar manipulations as in subsection \ref{planarize} (based on
equation \ref{thetashifts}), we put equation (\ref{axx}) in the
following form
\begin{eqnarray}
Z_{\sigma_L,\sigma_R }&=&k  \int_{-\infty}^{+\infty} ds_1 ds_2
\epsilon( \sigma_L) \epsilon( \sigma_R) \times \\
\nonumber & &
f_{\sigma_L}(s_1\tau+s_2,\tau)f_{\sigma_R}^{*}(s_1\tau+s_2,\tau)
e^{-\frac{\pi k}{\tau_2} |s_1 \tau+s_2|^2}~,
\end{eqnarray}
where $\epsilon=1$ for $\sigma=NS,\tilde R$, while $\epsilon=-1$ in the $\tilde{NS},R$ sectors.
This expression diverges when the holonomies take the values
\begin{eqnarray}
s_1 \tau + s_2 &=& m_1 \tau + m_2 \, .
\label{poles}
\end{eqnarray}
To deal with these infinities,
we cut out from the $x_1+ix_2 = s_1\tau+s_2$ plane, disks
 of radius $\varepsilon$ around each of these divergent contributions;
see figure \ref{cutdisk}.
This is
 modular invariant, since under modular transformations each disk transforms into another cutout.
This technique contrasts with previous regularization schemes that 
break modular invariance. Indeed, the region excised in
\cite{Hanany:2002ev} and all follow-ups is a sliver along one side of
the torus. It therefore transforms non-trivially under modular
transformations.  It does have the advantage of giving rise to a
strict implementation of the coset Gauss law constraint. Our
observations illustrate a new type of conundrum. Indeed, in
regularizing, we have the choice between giving up modular invariance
(or large diffeomorphism symmetry on the torus) or the Gauss law
constraint.

After introducing the regularization by cutting out little disks, we
find that the divergence at the origin gives rise to the contribution
\be\label{origin}
Z_{origin} = \frac{k}{\tau_2} \log(\varepsilon)
 \frac{|(\theta_3^4-\theta_4^4-\theta_2^4)|^2}{|\eta|^{12}}~,
\ee
which is modular invariant.
 It is the expected leading term one gets when
 considering a target space with a negligible curvature.  The
 divergence 
is due to states in the continuum
 with zero winding that propagate along the semi-infinite
 cylinder. The target-space details of the cut-off
are not clear, but very roughly speaking, $\varepsilon$ cuts off the size of the
 semi-infinite cylinder, a distance of order $\sqrt{k}\log(\varepsilon)$ from the tip.
 Corrections to the contribution (\ref{origin}) that scale like the volume of the
 cigar are obtained through evaluating the other divergences at the non-zero poles
in equation (\ref{poles}),
 that yield terms that
 scale like, schematically,
%
\be\label{wneq}
Z_{w}\sim k\log(\varepsilon) e^{- \pi\tau_2 k w^2 }.  \ee
These correspond to states in the continuous representations with
 $w\neq 0$, that propagate along the semi-infinite cylinder.

One question is whether there are corrections to the leading terms,
associated to the discrete states.
There are indeed correction terms that do not scale like the volume
(of the cigar, or the cap of the cigar).  These terms are found by expanding
the partition function in $\tau_2/k$ around the origin.
The leading
correction would seem to be
\be\label{uy}
Z_1 = \frac{1}{2 \pi} \frac{1}{|\eta|^{4}}
\partial_{\tau} \left( \frac{| \theta_3^4 -
    \theta_4^4-\theta_2^4|^2}{|\eta|^8} \right) \int d^2x \frac{(x_1+i
  x_2)^2}{|x_1+i x_2|} + c.c \, .  \ee
 Ignoring for a moment the
integral over $x_1$ and $x_2$, this term is exactly the non-modular
invariant contribution of the discrete states in the large $k$ limit
found in \cite{Giveon:2013ica} (see equation (5.17) in that paper and the relation
between the two in appendix \ref{linktoGI2}).  The contribution in equation (\ref{uy}),
however, vanishes due to the angular integration. It is thereby trivially
modular invariant. 

We encounter a puzzling situation. The discrete states make a distinct
contribution to the elliptic genus, but they appear to be well hidden
in the partition function expression.  Despite the vanishing of this
particular subleading term, there are two ways to see the imprint of
the discrete states on the partition function. The first way is to
calculate the next term in the expansion of the partition function.
The second way is to turn on R-charge chemical potential $\alpha$
and/or angular momentum chemical potential $\beta$. We hope to return to these
avenues in the near future.

\section{A UV/IR relation}
\label{uvir} {}From our results in section \ref{saddle}, it is clear
that the relationship between the holonomy space $(x_1, x_2)$ and
the target space is an ultraviolet/infrared map. Below, we discuss this
relation and use it in order to locate the discrete states in the
target space. This is an alternative to the wave-function analysis of
subsection \ref{location}, to
pinpoint where these states live.  We show that the UV/IR relation supports our
conclusion that these states are located at the string scale tip of the
cigar.

The fact that there is a UV/IR relation between the holonomy space and
the target space is most easily seen from the divergences we
encountered in the partition functions. In the holonomy space, we cut
out a small disk of radius $\varepsilon$ around each pole of the integrand,
which roughly
corresponds in the target space to a cutoff on the
semi-infinite cylinder (see the figures).
The smaller the cut-off $\varepsilon$ is, the larger the
part of the volume of the cylinder  that we take into account is. This illustrates the UV/IR nature of the
relationship between the holonomy space and the target space. Note
that the little disk need not be around the origin for this to be true, as is clear
from the discussion about the non-zero winding sector (around equation (\ref{wneq})).
In a sense, this UV/IR relation is like a Fourier transformation.

We wish to take advantage of this UV/IR relation in order to revisit the location
of the discrete states in the target space. It is clear they do not
live in the semi-infinite cylinder. The question is whether they are
located at the tip or the whole cap of the cigar.

With this motivation in mind, let us consider the elliptic genus. It has two
pieces. The non-holomorphic term is due to the difference in spectral densities
$\Delta\rho$
in the continuum. As emphasized in section \ref{generalities},
the spectral asymmetry $\Delta\rho$ is fixed by the asymptotic right-moving supercharge.
The asymptotic region is 
described by supergravity and so the relevant distance from the tip is
$\sqrt{k} l_s$, which is the only scale available at the supergravity
level. In the holonomy space, we saw that the contribution to the
non-holomorphic piece comes from a region of size $\sim
1/\sqrt{k}$. We find, therefore, that in string units
\be \frac{1}{\sqrt{k}} ~~\mbox{in holonomy space} ~~\Leftrightarrow ~~
\sqrt{k} ~~ \mbox{in target space}.  \ee 
The holomorphic piece in the
\eg and in the partition function originate in a region of size $\sim
1$ in the holonomy space. The UV/IR relation thus implies that they
are located, in target space, at a distance scale from the tip which is
a scale smaller than $\sqrt{k}$. The only other scale in the problem is the string
scale. This suggests that
\be
1 ~~\mbox{in holonomy space}  ~~\Leftrightarrow ~~ 1~~ \mbox{in target space},
\ee
and that the discrete states live at the tip of the cigar.

\subsubsection*{Twisted sectors as a ruler}
To make our argument sharper still, we consider the \eg of the cigar orbifolded
by a  $\mathbb{Z}_M$ subgroup of the R-symmetry group. We take $k\to\infty$ while keeping the order $M$
finite.  Now, there are also twisted sectors in the theory. These can
be used as a ruler that relates distances in the target space to
distances in holonomy space.  These twisted
degrees of freedom live a stringy distance away from the tip. To
determine their size in holonomy space, we calculate the \eg in this
case. The untwisted and twisted sectors are parameterized by two integers, $ p_1,
p_2 = 0,1,2...,M-1$, and their contribution to the \eg is
\begin{eqnarray}
\chi_{cos}^M(\tau,\alpha) &=& \frac{k}{M} \sum_{p_1,p_2 \in \mathbb{M}}
\int_0^1 d s_{1,2} \sum_{m,w \in \mathbb{Z}}
\frac{\theta_{11} (s_1 \tau+s_2 - \frac{k+1}{k}\alpha,\tau)}{\theta_{11}(s_1\tau+s_2
- \frac{\alpha}{k},\tau)} \nonumber \\
& & \times
e^{2 \pi i \alpha w} e^{ - \frac{k \pi}{\tau_2} | (m+\frac{p_1}{M}+s_2)+(w+\frac{p_2}{M}+s_1) \tau|^2} \, .
\end{eqnarray}
We can again rewrite this as an integral over the whole complex plane,
\begin{eqnarray}
& & \chi_{cos}^M(\tau,\alpha) = \\
& &  \frac{1}{M} \sum_{p_1,p_2 \in \mathbb{M}}
\int d^2 x
\frac{\theta_{11} (\sqrt{\frac{\tau_2}{k}} (x_1+ix_2) -
\frac{k+1}{k}\alpha,\tau)}{\theta_{11}(\sqrt{\frac{\tau_2}{k}} (x_1+ix_2)
- \frac{\alpha}{k},\tau)}
 e^{ - \pi  (x_1+\frac{p_1}{M} \sqrt{\frac{k}{\tau_2}} +\frac{p_2}{M} \sqrt{\frac{k}{\tau_2}} \tau_1)^2
+(x_2+\frac{p_2}{M}\sqrt{k\tau_2})^2)}, \nonumber
\end{eqnarray}
from which we see that the twisted sector
contribution is coming from a region of size $\sim 1$, located around
$(x_1, x_2)=-(\frac{p_1}{M} \sqrt{\frac{k}{\tau_2}} +\frac{p_2}{M} \sqrt{\frac{k}{\tau_2}} \tau_1, \frac{p_2}{M}\sqrt{k\tau_2})$.

We see that, indeed, regions of size of order $1$ in holonomy
space correspond to the tip of the cigar. Again, this supports our claim that
the discrete states also live at the tip of the cigar.

\section{The trumpet and  universality}
\label{trumpet}
In this section, we clarify a few aspects of the universality
discussed in the previous sections.  Note that the Wick rotation of the region
beyond the singularity of the $SL(2, \mathbb{R})_k/U(1)$ black hole
yields the trumpet geometry; it has an exact conformal field theory
description in terms of a $\mathbb{Z}_k$ orbifold of the cigar (see
e.g.~\cite{Giveon:1994fu} for a review). We study its elliptic genus in
the large level limit.

The elliptic genus of the trumpet is
\begin{eqnarray} 
\chi_{orb} &=& \chi_{orb}^{hol} + \chi_{orb}^{rem}~,
\end{eqnarray}
with~\cite{Troost:2010ud}
\begin{eqnarray}
\chi_{orb}^{hol} &=& \frac{i \theta_{11}(z,q)}{\eta^3}
\sum_{m \in \mathbb{Z}}
\frac{q^{k m^2} z^{2 m}}{1-z^{\frac{1}{k}} q^{m}}~,
\nonumber \\
\chi_{orb}^{rem} &=& -\frac{i \theta_{11} (z,q)}{\pi \eta^3} \sum_{p,w}
\int_{- \infty - i \epsilon}^{+ \infty - i \epsilon}
 \frac{ds}{2is-p+kw} q^{ \frac{s^2}{k} + \frac{(kw+p)^2}{4k}} z^{-\frac{kw+p}{k}}
\bar{q}^{ \frac{s^2}{k} + \frac{(kw-p)^2}{4k}}
\, .
\end{eqnarray}
Following the logic of section \ref{universality},
we focus first on the non-holomorphic piece,
from which one may determine the holomorphic piece via universality and modularity.
In the large $k$ limit,
the non-holomorphic piece does not depend on the level $k$; it reads
\begin{eqnarray}\label{nhtr}
\chi_{orb}^{non-hol}(k=\infty)=-\frac{i \theta_{11}}{ \eta^3}
\frac{i\alpha}{2\tau_2}~.
\end{eqnarray}
In particular, it is minus the non-holomorphic piece of the cigar. It
is also determined by the asymptotic supersymmetry and
cylindrical topology. For the trumpet's throat,
the sign of the spectral asymmetry $\Delta\rho$ is opposite to that of the cigar.

In section \ref{universality},
we used modularity as well as the assumption of a smooth filling in of the geometry
 to argue that the holomorphic piece is universal as well.
Now, the trumpet background is not in the same universality class as the cigar;
the topology is different,
and the string coupling diverges at the singularity.
In its weakly coupled T-dual description,
in terms of the $\mathbb{Z}_k$ orbifold of the cigar,
it has an orbifold fixed point at the tip. Hence,
the near tip background  behaves like a $\mathbb{Z}_k$ orbifold of $\mathbb{R}^2$,
and so its Witten index is $k$ (which contrasts with the cigar with Witten index $1$). The orbifold
singularity thus sets a new interior boundary condition for the allowed filling-in.
It  implies that the \eg 
scales like the level $k$.
Indeed, the holomorphic piece in the large level limit is
\be\label{31}
-\frac{i \theta_{11}}{ \eta^3}\frac{k}{2 \pi i\alpha}~.
\ee
This expression is modular covariant by itself for a theory with central charge $c=3$.
However, the central charge here is $c=3+6/k$.
The difference, $6/k$, goes to zero, but since the result (\ref{31}) scales like $k$,
this difference has a finite effect on the modular covariance properties of (\ref{31})
(see equation (\ref{tra})), even when $k\to\infty$.
This finite effect is exactly canceled against that of the
non-holomorphic piece (\ref{nhtr}). That is,
\begin{eqnarray}\label{agorb}
\chi_{orb}(k=\infty) &=&-\frac{i \theta_{11}}{ \eta^3}
\left(\frac{k}{2 \pi i\alpha}+\frac{i\alpha}{2\tau_2}\right)
\end{eqnarray}
is modular covariant with $c=3+6/k$.

A potentially interesting way to think about the diverging order $k$ contribution to the elliptic
genus of the trumpet, is to add to it the elliptic genus corresponding
to the Wick rotation of the {\it interior} of the black hole --
the regime between its horizon and singularity -- which is equal to~\cite{Giveon:2013hsa}
\be
\chi_{interior}=-\chi_{MM}(-k)~,
\ee
where
\begin{eqnarray}
\chi_{MM}(k) &=& \frac{\theta_{11}(z^{1-\frac{1}{k}},q)}{\theta_{11} (z^{\frac{1}{k}},q)}
\end{eqnarray}
is the elliptic genus of the $SU(2)_k/U(1)$ superconformal field theory -- an $N=2$ minimal model at level $k$.
The curious sum rule found in~\cite{Sugawara:2013hma,Giveon:2013hsa},
\be
\chi_{orb}+\chi_{interior}+\chi_{cigar}=0~,
\ee
which is correct for {\it any} positive integer level $k$,
implies that the interior of the black hole provides
an intriguing regularization of the regime beyond the singularity,
since the sum of their elliptic genera is indeed (the minus of) the elliptic genus of the cigar.
It would be very interesting to see if this holds more generally and, in particular,
for Schwarzschild black holes.

\section{Summary and discussion}
\label{conclusions}

In this work, we continued the investigation of perturbative string
theory in the background of large Euclidean black holes.  The main
focus in this paper was to study the \eg in the
$SL(2,\mathbb{R})_k/U(1)$ superconformal field theory in the large level limit and explore
its physical meaning.  We found that it is finite and contains two
pieces.  One is holomorphic and is due to the discrete states, the
other is non-holomorphic and is due to a mismatch in the densities of
worldsheet bosons and fermions in the continuum.
This mismatch is a feature that appears already at the gravity level and
is fully fixed by the topology of the cigar. In other words, it cannot
be avoided in any theory of quantum gravity.  Modularity of string
theory forces the existence of the holomorphic piece, and hence the
existence of the discrete states even for parametrically small
curvature.

Although the elliptic genus was analyzed for the {\it exact} superconformal
background of the two-dimensional black hole, universality and
modularity, discussed in this work, indicate that the properties above
are generic, including for perturbative strings propagating e.g. in
the background of a large four-dimensional Schwarzschild black hole.

We addressed the question: ``Where in target space do these discrete
states live in the large level limit?'' in two ways.
The answer is that they live at a stringy distance from the tip
of the cigar.  This, we view, as evidence for the claims made
in~\cite{Giveon:2012kp,Giveon:2013ica} that the tip of the cigar is special in string theory
even for parametrically small curvature. There are degrees of freedom in the theory
with definite conformal dimension and a wave function that is
supported at a stringy distance from the tip. In contrast, in the free scalar conformal field
theory associated to $\mathbb{R}^2$, states with definite conformal dimension are smeared over the whole
plane.

In short, the \eg has enough structure to indicate that perturbative
string theory in the background of a large black hole is yet to reveal
its full peculiarities.
Questions remain, like: 
``What are the implications, if any, of the degrees
of freedom at the tip for an infalling observer?''  This, we believe,
should be determined by the Wick rotation of the interaction between
the tip degrees of freedom and the gravitons, dilaton, et cetera, in the
limit where the curvature is small. We hope to make progress on that
front as well.

\subsection*{ Acknowledgments}
We thank Ofer Aharony, Sujay Ashok, Sameer Murthy and Suresh Nampuri for discussions.
J.T. would like to thank the
Hebrew University of Jerusalem as well as Tel-Aviv University for
their warm hospitality during his visit.  The work of A.G. and N.I. is
supported in part by the I-CORE Program of the Planning and Budgeting
Committee and the Israel Science Foundation (Center No. 1937/12). The
work of A.G. is supported in part by the BSF -- American-Israel
Bi-National Science Foundation, and by a center of excellence
supported by the Israel Science Foundation (grant number 1665/10).

\appendix

\section{Useful formulas}
In this appendix, we collect various formulas that we put to use in the main
body of the paper.

\subsection{Conventions and properties of theta functions}
Our conventions for theta functions as well as some of the properties
of their derivatives, relations to series, et cetera are recalled in the
following.
\subsubsection{Product and sum formulas}
 We use the product and sum formulas for theta functions as well as the eta function:
\begin{eqnarray}
\eta (q) &=& q^{ \frac{1}{24}} \prod_{m=1}^\infty (1-q^m)
\nonumber \\
\theta_{00} (z,q) &=&  \prod_{m=1}^\infty (1-q^m)(1+z q^{m-1/2})(1+z^{-1} q^{m-1/2})
= \sum_{n \in \mathbb{Z}} q^{ \frac{n^2}{2}} z^n
\nonumber \\
&=& \theta_3(z,q) = \theta[0,0](z,q)
 \\
\theta_{01} (z,q) &=&  \prod_{m=1}^\infty (1-q^m)(1-z q^{m-1/2})(1-z^{-1} q^{m-1/2})=
\sum_{n \in \mathbb{Z}} (-1)^n q^{ \frac{n^2}{2}} z^n \nonumber \\
&=& \theta_4(z,q) = \theta[0,1/2](z,q)
\nonumber\\
\theta_{10} (z,q) &=& (z^{\frac{1}{2}}+z^{-\frac{1}{2}}) q^{\frac{1}{8}}
\prod_{m=1}^\infty (1-q^m)(1+z q^m)(1+z^{-1} q^m)= \sum_{n \in \mathbb{Z}} q^{ \frac{(n-\frac{1}{2})^2}{2}} z^{n-\frac{1}{2}}
 \nonumber \\
&=& \theta_2(z,q) = \theta[1/2,0](z,q)
\nonumber \\
\theta_{11} (z,q) &=& i (z^{\frac{1}{2}}-z^{-\frac{1}{2}})
q^{\frac{1}{8}} \prod_{m=1}^\infty (1-q^m)(1-z q^m)(1-z^{-1} q^m)= -i \sum_{n \in \mathbb{Z}}
(-1)^n q^{ \frac{(n-\frac{1}{2})^2}{2}} z^{n-\frac{1}{2}} \nonumber \\
&=& -\theta_1(z,q) = \theta[1/2,1/2](z,q) \nonumber
\, ,
\end{eqnarray}
where $q=e^{2 \pi i \tau}$ and $z= e^{2 \pi i \alpha}$.
\subsubsection{Properties}
We have the properties:
\begin{eqnarray}
\theta_1(0,\tau) &=& 0
\nonumber \\
\partial_\tau \theta_1 (0,\tau) &=& 0
\nonumber \\
\partial_\alpha \theta_{\sigma \neq 1} (0,\tau) &=& 0
 \\
4 \pi i \partial_\tau \theta_{\sigma} &=& \partial_\alpha^2 \theta_{\sigma}
\label{heat}
\end{eqnarray}
\subsubsection{Shifts}
We can shift arguments of $\theta$ functions using the formula:
\begin{eqnarray}
\theta[a,b] (\alpha + m_2 + m_1 \tau,\tau)
&=& (-1)^{2 a m_2} (-1)^{2 b m_1} z^{-m_1} q^{ - \frac{m_1^2}{2}}
\theta[a,b] (\alpha,\tau) \, \label{thetashifts}
\end{eqnarray}
in a notation where $a,b$ are zero or one half.

%

\subsubsection{Derivatives}
We can use the product formulae for the theta functions to obtain their
derivatives.
Their $\alpha$-derivatives are:
\begin{eqnarray}
\partial_\alpha \theta_{00} (z,q)
&=& \theta_{00} (z,q)\sum_{m=1}^\infty 2 \pi i ( \frac{z q^{m-1/2}}{1+z q^{m-1/2}}
- \frac{z^{-1} q^{m-1/2}}{1+z^{-1} q^{m-1/2}})
\nonumber \\
&=& \theta_{00}(z,q) (z-z^{-1}) \sum_{m=1}^\infty 2 \pi i  \frac{q^{m-1/2}
}{(1+z q^{m-1/2})(1+z^{-1} q^{m-1/2})}
\nonumber \\
\partial_\alpha \theta_{01} (z,q)
&=& \theta_{01}(z,q) \sum_{m=1}^\infty 2 \pi i ( \frac{-z q^{m-1/2}}{1-z q^{m-1/2}}
+ \frac{z^{-1} q^{m-1/2}}{1-z^{-1} q^{m-1/2}})
\nonumber \\
&=& \theta_{01}(z,q)  (z^{-1}-z) \sum_{m=1}^\infty 2 \pi i  \frac{q^{m-1/2}}{(1-z q^{m-1/2})
(1-z^{-1} q^{m-1/2})}
\nonumber \\
\partial_\alpha \theta_{10}(z,q)
  &=& \theta_{10}(z,q)  (
\pi i \frac{z^{\frac{1}{2}}-z^{-\frac{1}{2}}}{z^{\frac{1}{2}}+z^{-\frac{1}{2}}} + \sum_{n=1}^\infty 2 \pi i (  \frac{z q^n}{1+zq^n} - \frac{z^{-1} q^n}{1+z^{-1} q^n}))
\nonumber \\
& =&   \theta_{10}(z,q) 2 \pi i ( \frac{1}{2} \frac{1-z}{z+1}
+ (z-z^{-1}) \sum_{m=1}^\infty  \frac{q^{n}}{(1+z q^n)(1+z^{-1} q^n)} )
\nonumber \\
  \partial_\alpha \theta_{11}(z,q)
  &=& \theta_{11}(z,q)  ( \pi i \frac{z^{\frac{1}{2}}+z^{-\frac{1}{2}}}{z^{\frac{1}{2}}-z^{-\frac{1}{2}}} + \sum_{n=1}^\infty 2 \pi i ( - \frac{z q^n}{1-zq^n} + \frac{z^{-1} q^n}{1-z^{-1} q^n}))
\nonumber \\
& =&   \theta_{11}(z,q) 2 \pi i ( \frac{1}{2} \frac{1+z}{z-1}
+ (z^{-1}-z) \sum_{m=1}^\infty  \frac{q^{n}}{(1-z q^n)(1-z^{-1} q^n)} )
\label{series}
\end{eqnarray}

\subsection{Further properties applied}
Particular applications of these properties, as well as further
formulas of use are the following.
\subsubsection{Infinite series, theta functions and a partition sum}
\label{linktoGI2}
In \cite{Giveon:2013ica}, a contribution to the partition function was identified
proportional to various infinite series (related to sums of odd divisors). Those
series can be rewritten in terms of derivatives of $\theta$ functions as follows:
\begin{eqnarray}
\Sigma_R &=& \sum_{n=0}^\infty \frac{n q^n}{1+q^n}
= \frac{1}{48}  (\theta_3(q)^4 + \theta_4(q)^4-1)
= \frac{1}{2} q \partial_q \log \frac{ q^{- \frac{1}{12}} \theta_2(q) }{\eta}
\nonumber \\
\Sigma_{NS} &=& \sum_{n=0}^\infty (2n+1) \frac{q^{n+1/2}}{1+q^{n+1/2}}
= \frac{\theta_2(q)^4-\theta_4(q)^4+1}{24}
= q \partial_q  \log \frac{q^{\frac{1}{24}} \theta_3}{\eta}
\nonumber \\
\Sigma_{\tilde{NS}} &=&
 \sum_{n=0}^\infty (2n+1) \frac{q^{n+1/2}}{1-q^{n+1/2}}
= \frac{\theta_2(q)^4 + \theta_3(q)^4 -1}{24}
= -q \partial_q  \log \frac{q^{\frac{1}{24}} \theta_4}{\eta}\, .
\end{eqnarray}
These equations are the main ingredients in putting the result of \cite{Giveon:2013ica}
in the form of equation (\ref{uy}).
We also made use of the relation:
\begin{eqnarray}
\eta(q) &=& ( \frac{\partial_\alpha \theta_{11} (0,\tau)}{- 2 \pi})^{\frac{1}{3}}
\end{eqnarray}
as well as:
\begin{eqnarray}
- 2 \pi i \partial_\tau \log \eta &=& \frac{\pi^2}{6} E_2
\label{logeta} 
\end{eqnarray}
where $E_2$ is the second Einsenstein series, whose modular completion is
\begin{eqnarray}
\hat{E}_2 &=& E_2 - \frac{3}{\pi \tau_2}
\end{eqnarray}
which imply that:
\begin{eqnarray}
\partial_\tau \log ( \tau_2 |\eta|^4) &=& \frac{1}{2 i \tau} + \frac{i \pi}{6} E_2
= \frac{i \pi}{6} \hat{E}_2 \, .
\end{eqnarray}
\subsubsection{The expansion of the large level genus near $\alpha=0$}
The link between the second Eisenstein series and the derivative
of theta functions (near $\alpha=0$) can be found either through the manipulations:
\begin{eqnarray}
E_2 (\tau) &=& 1 -24 \sum_{n=1}^\infty \frac{n q^n}{1-q^n}
=1-24 \sum_{n=1}^\infty \sum_{m=0}^\infty n q^{n(m+1)}
=1-24 \sum_{m=1}^\infty \frac{q^m}{ (1-q^m)^2}
\end{eqnarray}
and observing that this series appears in equation (\ref{series}), or via
expansion near $\alpha=0$, the heat equation (\ref{heat}) as well as equation (\ref{logeta}).
Using these formulas, we can for instance describe the second term in the expansion
of the large level limit of the cigar elliptic genus in terms of $\alpha$.
Near $\alpha=0$, the cigar elliptic genus contains the terms:
\begin{eqnarray}
\chi(k = \infty)
& \approx &
\frac{\theta_{11}(\alpha,\tau)}{\eta^3(q)}
( - \frac{1}{2 \pi \alpha} + \alpha \frac{\pi}{6} \hat{E}_2)
\approx
 1 - \alpha^2  \frac{\pi^2}{2} E_2 + \frac{\pi \alpha^2}{\tau_2} \, .
\end{eqnarray} 
It can be shown that this expression has the right modular covariance near $\alpha=0$,
using the transformation rule for the elliptic genus as well as the (modularly completed) second
Eisenstein series $\hat{E}_2$.

\end{document}